\documentclass{article}
\usepackage{spconf,amsmath,graphicx,makecell,dblfloatfix}
\usepackage{subfig}

\title{DOUBLE BINNABLE RGB, RGBW AND LMS COLOR FILTER ARRAYS}
\name{Mritunjay Singh, Tripurari Singh}
\address{Image Algorithmics \\
  Bethesda, MD 20816 \\
USA}
\begin{document}
%
\maketitle
\begin{abstract}
  New, high-resolution CMOS image sensors for mobile phones have moved
  beyond the single-binnable Quad Bayer and RGBW-Kodak patterns to the
  double binnable Hexadeca Bayer pattern featuring 4x4 tiles of
  like-colored pixels. Pixel binning enables high-speed, low-power
  readout in low-resolution modes and, more importantly, a reduction
  of read noise via floating diffusion binning.
  
  In this paper, we present the non-intuitive result that Nona and
  Hexadeca Bayer can be superior to Quad Bayer in demosaicking quality
  due to degeneracies in the latter's spectrum. However, Hexadeca
  Bayer suffers from the weakness of generating Quad Bayer after one
  round of binning.

  We introduce novel double binnable RGBW and LMS CFAs, composed of
  2x2 tiles capable of 4:1 floating diffusion binning, that are free
  from spectral degeneracies and thus demosaic well in full resolution
  and both binned modes. RGBW offers a 6 dB low-light, read
  noise-limited, SNR advantage over Quad and Hexadeca Bayer in all
  resolution modes, while for LMS, the corresponding advantage is 4.2
  dB.
\end{abstract}
\begin{keywords}
  Binning, Floating Diffusion Binning, Charge Domain Binnning, Color
  Filter Array, Hexadeca Bayer, RGBW, RGBC, LMS.
\end{keywords}
\section{Introduction}
\label{sec:intro}

The rapid shrinkage of mobile CMOS image sensors' pixel pitch,
triggered by advancements in stacked die technology, has resulted in
sensors with a very large number of low SNR pixels, leading to the
popularity of pixel binning \cite{barna2013method}. In addition to the
obvious benefits of high-speed, low-power readouts in low-resolution
modes, pixel binning also helps reduce read noise by accumulating
charge from the binned pixels in their common floating diffusion - a
noiseless process - before encountering the noisy source follower and
the downstream read circuit \cite{chu2006improving}.

Binning of 4 pixels to 1 results in a 6 dB SNR improvement in
low-light, read noise limited settings if the pixel values are summed
in the voltage or digital domains. This is comparable to the SNR
improvement expected by reading out the full-resolution image followed
by downsizing to half resolution. Floating diffusion binning, on the
other hand, can deliver a 12 dB SNR improvement in low-light, read
noise-limited settings by replacing 4 trips through the noisy source
follower for every binned pixel read with 1. In bright light, photon
shot noise-limited settings, all binning methods provide similar SNR
improvement, which is also broadly similar to that obtained by reading
out the full-resolution image followed by downsizing.

Floating diffusion binning is only possible for pixels that share a
readout circuit. Tiles of 2x2 pixels share a readout circuit in most
high-resolution CMOS sensors, limiting floating diffusion binning to
4:1 binning. Variants include a mode that restricts floating diffusion
binning to 2:1 to avoid integration times short enough to interfere
with LED flicker. Another variant uses a switchable bank of floating
diffusions to provide variable capacitance - or conversion gain - in
order to avoid saturation and extend the dynamic range of the pixel
\cite{venezia20181,park20190}.

\subsection{Popular Binnable Color Filter Arrays}

The most popular binnable CFA is the Quad Bayer pattern, shown in
Figure \ref{fig:quadBayer}, which replaces each pixel of the Bayer
pattern with a 2x2 tile of 4 pixels of the same color. Extensions of
the Quad Bayer pattern are the newer Nona Bayer and Hexadeca Bayer
patterns that replace each Bayer pixel with a tile of 3x3 and 4x4
pixels, respectively. The Hexadeca Bayer pattern allows for two rounds
of binning, generating the Quad Bayer pattern after the first round of
binning and the Bayer pattern after the second round.

A much less popular CFA is the RGBW-Kodak pattern \cite{kumar2009new},
see Figure \ref{fig:rgbwKodak}, which allows 2:1 pixel binning along
the diagonal direction. While the RGBW-Kodak CFA yields a 6 dB SNR
improvement over Quad Bayer in full resolution, low-light read
noise-limited settings, the problem is that full-resolution mode is
typically not used in low light. Noise obscures fine detail in low
light, making the lower resolution, higher SNR, floating diffusion
binned images visually superior - even after upscaling to full
resolution. The SNR lead of RGBW-Kodak over Quad Bayer shrinks to 3 dB
after one round of binning since the former is limited to 2:1 floating
diffusion binning while the latter is capable of 4:1 floating
diffusion binning. Furthermore, the RGBW-Kodak requires the readout of
twice as much data in the binned mode as Quad Bayer. The combination
of RGBW-Kodak's modest 3 dB SNR improvement over Quad Bayer at the
expense of higher power consumption by both the sensor's readout
circuit and the demosaicker has hampered its prospects in the mobile
market.

\section{Spectral Analysis of Color Filter Arrays}

Color Filter Arrays can be analyzed in the frequency domain according
to the method of \cite{LukacBook} which allows the CFA to be
represented as the sum of modulated chrominance signals and a baseband
luminance signal.

In the spectrum of each CFA analyzed in the following sections, the
luminance and chrominance signals are plotted as circles since the
resolution of the camera lens is assumed to be equal in all
directions. Furthermore, the radius of each chrominance band is
plotted as half that of the luminance band in keeping with the ``half
bandwidth chrominance'' assumption of most demosaickers. This
assumption is fair in most cases due to the high correlation between
the R, G, and B color planes. Furthermore, adaptive demosaickers
assume both luminance and chrominance bandwidths to be locally low
along the direction of edges - a fact not depicted in the following
figures.

\section{The Bayer Family}

\begin{figure}[h]
\begin{minipage}[b]{.48\linewidth}
  \centering
  \centerline{\includegraphics[width=4.0cm]{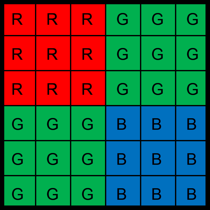}}
\end{minipage}
\hfill
\begin{minipage}[b]{0.48\linewidth}
  \centering
  \centerline{\includegraphics[width=4.0cm]{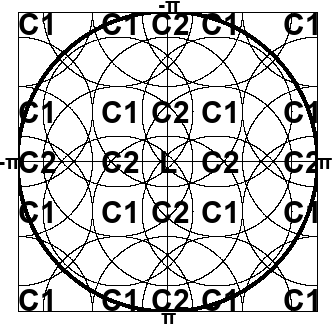}}
\end{minipage}
\caption{Nona Bayer CFA (left), and its spectrum (right).}
\label{fig:nonaBayer}
\end{figure}

\begin{figure}
\begin{minipage}[b]{.48\linewidth}
  \centering
  \centerline{\includegraphics[width=4.0cm]{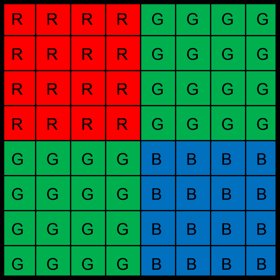}}
\end{minipage}
\hfill
\begin{minipage}[b]{0.48\linewidth}
  \centering
  \centerline{\includegraphics[width=4.0cm]{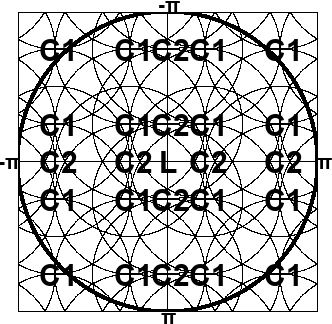}}
\end{minipage}
\caption{Hexadeca Bayer CFA (left), and its spectrum (right).}
\label{fig:hexaDecaBayer}
\end{figure}

The Quad Bayer, Nona Bayer and Hexadeca Bayer CFAs and their spectra
are shown in Figures \ref{fig:nonaBayer}, \ref{fig:hexaDecaBayer} and
\ref{fig:quadBayer}. In each case, the luminance $l=R+2G+B$, and the
chrominances $c_1=2G-R-B$ and $c_2=R-B$.

\subsection{Nona and Hexadeca Bayer}

Nona \cite{oh20200} and Hexadeca Bayer spectra contain a larger number
of chrominance signals modulated at various carrier frequencies, some
of which are close to each other or to the luminance. The large number
of chrominance signals themselves is not a problem since, along with
the luminance, they have only three degrees of freedom - that of red,
green, and blue.

The low separation and the resulting overlap of the chrominance
spectra is not a problem either. As \cite{singh2011icip} shows,
spectral overlap can be disregarded except under certain degenerate
conditions. A simple necessary, though not sufficient, condition for
good recovery of chrominance signals is for at least one copy of the
linearly independent chrominance signals to be sufficiently separated
from each other and from the luminance.

Both Nona and Hexadeca Bayer can be demosaicked with good resolution
and low false color using the technique of \cite{singh2011icip}, with
the residual false color being amenable to removal by post processing.

\subsection{The Unfortunate Case of Quad Bayer}

\begin{figure}[h]
\begin{minipage}[b]{.48\linewidth}
  \centering
  \centerline{\includegraphics[width=4.0cm]{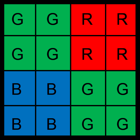}}
\end{minipage}
\hfill
\begin{minipage}[b]{0.48\linewidth}
  \centering
  \centerline{\includegraphics[width=4.0cm]{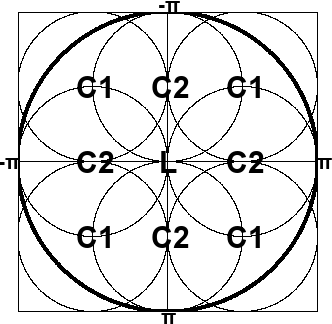}}
\end{minipage}
\caption{Quad Bayer CFA (left), and its spectrum (right).}
\label{fig:quadBayer}
\end{figure}

The Quad Bayer spectra, shown in Figure \ref{fig:quadBayer}, has
carriers in the interior of the spectrum, located at $(\pm
\frac{\pi}{2}, 0)$, $(\pm \frac{\pi}{2}, \pm \frac{\pi}{2})$ and $(0,
\pm \frac{\pi}{2})$. The absence of any chrominance carriers
sufficiently removed from luminance makes it impossible for any
demosaicker that relies on the half chrominance bandwidth assumption
to perform well. Since most demosaickers, directly or indirectly, make
this assumption, they generate artifacts that can be traced to
luminance-chrominance confusion at the chrominance carrier
frequencies.

\section{Binnable RGBW CFAs}

\begin{figure}[h]
\begin{minipage}[b]{.48\linewidth}
  \centering
  \centerline{\includegraphics[width=4.0cm]{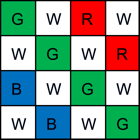}}
\end{minipage}
\hfill
\begin{minipage}[b]{0.48\linewidth}
  \centering
  \centerline{\includegraphics[width=4.0cm]{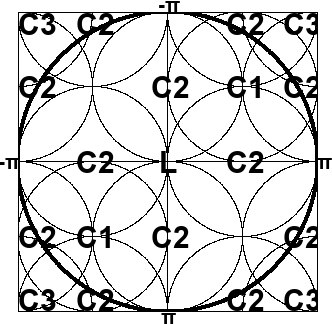}}
\end{minipage}
\caption{RGBW-Kodak CFA (left), and its spectrum (right).}
\label{fig:rgbwKodak}
\end{figure}

The only binnable RGBW CFA in production is the RGBW-Kodak pattern
\cite{kumar2009new} shown in Figure \ref{fig:rgbwKodak}. RGBW-Kodak
outputs 2 pixels for each 2x2 tile after binning, instead of the one
output by Quad Bayer. This results in two images: a Bayer mosaic and a
W color plane. The W color plane has the following uses:
\begin{enumerate}
\item the (Bayer mosaic minus W) difference image is easier to
  demosaic than the Bayer mosaic alone and results in better
  resolution and fewer artifacts
\item a fusion algorithm can use the higher SNR W color plane to clean
  up the noisier demosaicked RGB color planes.
\end{enumerate}
    
RGBW has higher SNR than RGB systems on image features with relatively
unsaturated colors owing to the greater radiant energy captured by W
pixels than either of R, G, or B. On highly saturated red, green and
blue features, W pixels do not capture much more energy than R, G, B
pixels, and the SNR improvement stems from the greater density of
pixels sensitive to the color in question. Since half the pixels are
W, 3/4 of all pixels are sensitive to G and 5/8 to R or B instead of
1/2 and 1/4 respectively for the Bayer family of CFAs.

The spectrum of full resolution RGBW-Kodak, shown in Figure
\ref{fig:rgbwKodak}, consists of the luminance $l=R+2G+B+4W$ and the
three chrominance signals $c_1=R-2G+B$, $c_2=R-B$, $c_3=R+2G+B-4W$
together accounting for the four degrees of freedom of R, G, B,
W. RGBW has been conjectured to have just three degrees of freedom
\cite{wang2011icip, rafinazari2015demosaicking}, but this has not been
borne out in practice as W is not a linear combination of R, G, B as
realized by practical color filters.

The problem with RGBW-Kodak is that chrominance $c_1 = R-2G+B$ has
only one copy and it is modulated at the relatively low carrier
frequency of $\pm(\frac{\pi}{2}, \frac{\pi}{2})$. This creates
confusion between luminance and chrominance resulting in substantial
false color in the presence of diagonal luminance features of
approximately $\pm(\frac{\pi}{2}, \frac{\pi}{2})$ frequency.

\subsection{A Novel Single Binnable RGBW CFA}

\begin{figure}[h]
\begin{minipage}[b]{.48\linewidth}
  \centering
  \centerline{\includegraphics[width=4.0cm]{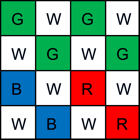}}
\end{minipage}
\hfill
\begin{minipage}[b]{0.48\linewidth}
  \centering
  \centerline{\includegraphics[width=4.0cm]{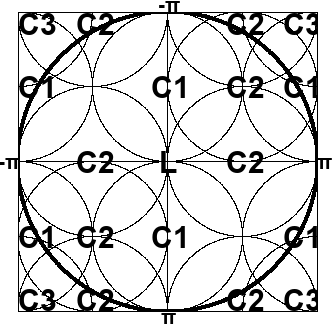}}
\end{minipage}
\caption{RGBW-IA CFA (left), and its spectrum (right).}
\label{fig:rgbwIA}
\end{figure}

We propose the novel RGBW-IA pattern shown in Figure \ref{fig:rgbwIA}
that redistributes RGBW-Kodak's 2x2 tiles so as to break up the green
diagonal and thus generate more chrominance carriers.

Its spectrum is also shown in Figure \ref{fig:rgbwIA}. The definition
of $l=R+2G+B+4W$, $c_1=R-2G+B$, $c_2=R-B$, $c_3=R+2G+B-4W$ are
identical to those of the RGBW-Kodak CFA. Having at least one copy of
$c_1$, $c_2$, $c_3$ removed from the luminance $l$ reduces crosstalk
with it; having multiple copies each of $c_1$, $c_2$ allows the
demosaicker to adaptively pick the cleaner copy depending on edge
orientation. The method of \cite{singh2011icip} reconstructs images
with high resolution and false color that is low enough to be
removable by post-processing. False color removal algorithms are a
topic of research in themselves and out of the scope of this paper.

\begin{figure}
  \begin{center}
    \includegraphics[width=2.0cm]{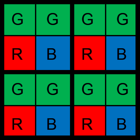}
        \includegraphics[width=2.0cm]{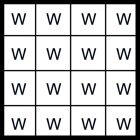}
  \end{center}
\caption{Binned RGBW-IA, tiled 4 times.}
\label{fig:stripedG}
\end{figure}

Like RGBW-Kodak, RGBW-IA can be binned by summing the pair of
diagonally placed W pixels and the pair of diagonally placed R, G or B
pixels in each 2x2 tile yielding an RGB mosaic and a W color
plane. Unlike RGBW-Kodak, the mosaic generated by binning RGBW-IA is
not the Bayer pattern; instead it is the less demosaicking friendly
pattern shown in Figure \ref{fig:stripedG}. However, the availability
of the W color plane allows the easy to demosaic (mosaic minus W)
color difference to be computed, demosaicked and the R, G, B color
planes to be reconstructed by adding back W. The SNR performance of
RGBW-IA is similar to RGBW-Kodak.

\subsection{A Novel Double Binnable RGBW CFA}

\begin{figure}[h]
\begin{minipage}[b]{.48\linewidth}
  \centering
  \centerline{\includegraphics[width=4.0cm]{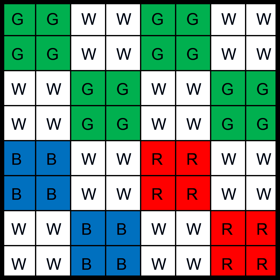}}
\end{minipage}
\hfill
\begin{minipage}[b]{0.48\linewidth}
  \centering
  \centerline{\includegraphics[width=4.0cm]{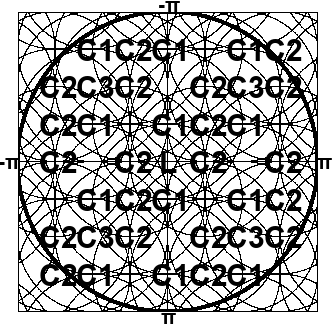}}
\end{minipage}
\caption{Quad-RGBW CFA (left), and its spectrum (right).}
\label{fig:quadIA}
\end{figure}

We generate the novel Quad-RGBW pattern, shown in Figure
\ref{fig:quadIA}, by replacing each pixel of the single binnable
RGBW-IA pattern with a 2x2 tile of pixels of the same color. The
result is a double binnable CFA with the first round supporting 4:1
floating diffusion binning.

The spectrum of the Quad-RGBW pattern, shown in Figure
\ref{fig:quadIA}, has the same definition of $l=R+2G+B+4W$,
$c_1=R-2G+B$, $c_2=R-B$, $c_3=R+2G+B-4W$ as the RGBW-IA and the
RGBW-Kodak CFAs. While the spectrum contains a large number of closely
spaced chrominance signals, they have at least one copy removed from
the luminance $l$ which reduces their crosstalk with
luminance. Multiple copies of $c_1$, $c_2$, $c_3$ allows the
demosaicker to adaptively pick the cleaner copies depending on edge
orientation. The method of \cite{singh2011icip} reconstructs images
with high resolution and false color that is low enough to be
removable by post-processing.

The first round of 4:1 diffusion binning gives Quad-RGBW a 3 dB low
light, read noise limited, SNR advantage over binned RGBW-Kodak. At
full resolution, and all bin modes, Quad-RGBW has a 6 dB low light,
read noise limited and 3 dB bright light, shot noise limited SNR
advantage over Quad Bayer and Hexadeca Bayer.

\section{Binnable LMS CFAs}

In this section we develop binnable versions of the LMS CFAs proposed
in \cite{LMScamera}. The spectral responses of the L, M, S pixels are
identical to those of the the (L)ong, (M)edium and (S)hort cones of
the human retina. Additionally, like the retina, the LMS CFA contains
a larger number of L, M pixels but fewer S pixels.

While M, S pixels are close analogs of the G, B pixels of Bayer
sensors, L is a shade of yellow. The spectral response of L is
responsible for the improved color accuracy as well has the SNR and
dynamic range advantage of LMS over Bayer.

\subsection{A Novel Single Binnable LMS CFA}

\begin{figure}
\begin{minipage}[b]{.48\linewidth}
  \centering
  \centerline{\includegraphics[width=4.0cm]{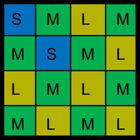}}
\end{minipage}
\hfill
\begin{minipage}[b]{0.48\linewidth}
  \centering
  \centerline{\includegraphics[width=4.0cm]{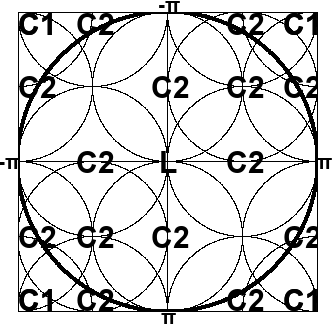}}
\end{minipage}
\caption{Single binnable LMS CFA (left), and its spectrum (right).}
\label{fig:LMSdiag}
\end{figure}

We propose the novel single-binnable LMS CFA shown in Figure
\ref{fig:LMSdiag}. The pattern is composed of pixel pairs of like
color arranged along diagonals thereby allowing for 2:1 diagonal
binning.

The spectrum of the single binnable LMS CFA consists of $l=3L+4M+S$,
$c_1=6L-8M+2S$, $c_2=L-S$. Both chrominances, $c_1$, $c_2$ have copies
near the periphery of the spectrum and sufficiently removed from
luminance to prevent crosstalk with it. Having multiple copies of
$c_2$ allows the demosaicker to adaptively pick the cleaner copy
depending on edge orientation. The method of \cite{singh2011icip}
demosaicks the single binnable LMS with high resolution and low false
color.

\begin{table*}
\begin{center}
  \begin{tabular}{| l | c | c | c | c | c |} 
    \hline
    & {\bf Kodim} & {\bf ISO 12233} & {\bf CZP} & {\bf TE42} & {\bf Newsprint} \\
        {\bf CFA} & {\bf Image Set} & {\bf Chart} & {\bf Chart} & {\bf Chart} & {\bf Image} \\ \Xhline{3\arrayrulewidth}
        {\bf Quad Bayer} & 38.4 & 44.4 & 38.6 & 44.1 & 37.5 \\ \hline
        {\bf RGBW-Kodak} & 38.9 & 47.4 & 41.6 & 45.7 & 45.2 \\ \hline
        {\bf RGBW-IA} & {\bf 42.4} & {\bf 47.6} & {\bf 43.9} & {\bf 46.6} & 45.1 \\ \hline
        {\bf LMS} & 39.6 & 47.3 & 43.7 & 46.4 & {\bf 46.7} \\ \hline
  \end{tabular}
\end{center}
\caption{Demosaicking Performance of single binnable CFAs (PSNR in
  dB).}
\label{tab:singlepsnr}
\end{table*}

\begin{table*}
\begin{center}
  \begin{tabular}{| l | c | c | c | c | c |} 
    \hline
    & {\bf Kodim} & {\bf ISO 12233} & {\bf CZP} & {\bf TE42} & {\bf Newsprint} \\
        {\bf CFA} & {\bf Image Set} & {\bf Chart} & {\bf Chart} & {\bf Chart} & {\bf Image} \\ \Xhline{3\arrayrulewidth}
        {\bf Hexadeca Bayer} & 38.4 & 45.1 & 41.0 & 44.1 & 39.7 \\ \hline
        {\bf Quad-RGBW-IA} & 38.4 & 46.0 & 41.3 & 44.0 & {\bf 42.8} \\ \hline
        {\bf Quad LMS} & {\bf 39.8} & {\bf 46.1} & {\bf 44.2} & {\bf 44.5} & 42.4 \\ \hline
  \end{tabular}
\end{center}
\caption{Demosaicking Performance of double binnable CFAs (PSNR in
  dB).}
\label{tab:doublepsnr}
\end{table*}

\subsection{A Novel Double Binnable LMS CFA}

We generate the novel Quad-LMS pattern, shown in Figure
\ref{fig:quadIA}, by replacing each pixel of the single binnable LMS
pattern with a 2x2 tile of pixels of the same color. The result is a
double binnable CFA with the first round supporting 4:1 floating
diffusion binning.

\begin{figure}[h]
\begin{minipage}[b]{.48\linewidth}
  \centering
  \centerline{\includegraphics[width=4.0cm]{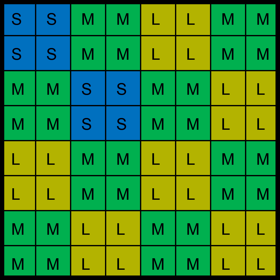}}
\end{minipage}
\hfill
\begin{minipage}[b]{0.48\linewidth}
  \centering
  \centerline{\includegraphics[width=4.0cm]{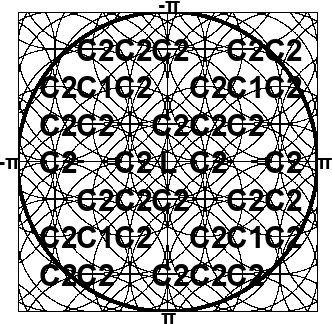}}
\end{minipage}
\caption{Double binnable LMS CFA (left), and its spectrum (right).}
\label{fig:quadLMSdiag}
\end{figure}

The spectrum of the Quad-LMS pattern, shown in Figure
\ref{fig:quadLMSdiag}, has the same definition of $l=3L+4M+S$,
$c_1=6L-8M+2S$, $c_2=L-S$ as the single binnable LMS CFA. While the
spectrum contains a large number of closely spaced chrominance
signals, they have at least one copy removed from the luminance $l$
which reduces their crosstalk with luminance. Multiple copies of
$c_1$, $c_2$ allows the demosaicker to adaptively pick the cleaner
copies depending on edge orientation. High quality demosaicking of
Quad LMS is possible with the algorithm of \cite{singh2011icip}.

\section{Experimental Results}

We conducted a simulation study to compare the performance of the CFAs
in the noiseless case. Starting with the sRGB ground truth images, we
linearized the images by reversing the sRGB tone map, applied a
diffraction-limited lens model with an Airy diameter of 2 pixels,
followed by conversion to the image sensor color space by applying the
inverse of the following typical mobile image sensor color correction
matrix:
\begin{center}
$ \begin{bmatrix}
    1.81 & -0.53 & -0.28 \\
    -0.30 & 1.38 & -0.08 \\
    -0.13 & -0.33 &1.46
  \end{bmatrix} $
\end{center}

Finally, we mosaicked the images with the CFA pattern in question to
generate the raw data.

Our processing pipeline consisted of demosaicking with the method of
\cite{singh2011icip}, followed by post-processing to remove false
color, and finally, chrominance denoising. Chrominance denoising in
the W, R-W, G-W, B-W color space is an essential step for an RGBW
system as it enables it to clean up R, G, B channels with the high SNR
W channel. While chrominance denoising is not an essential step for
some CFAs, we employed it in pipelines for all CFAs to ensure fair
comparisons.

We then converted to the sRGB color space and applied the sRGB tone
map. We did not perform luminance denoising or any other
post-processing.

We tested on Kodak's ``kodim'' image set and also the following test
charts popular in industry: TE42, ISO 12233 resolution target,
Circular Zone Plate. Finally, we stress-tested each CFAs ability to
capture fine, irregular features with a newsprint image.

We employed the PSNR metric to measure the demosaicking quality of the
CFAs. The results for single binnable CFAs are summarized in table
\ref{tab:singlepsnr} and those for double binnable CFAs are summarized
in table \ref{tab:doublepsnr}.

\begin{figure}[h]
\begin{minipage}[b]{\linewidth}
  \centering
  \centerline{\includegraphics[width=8.5cm]{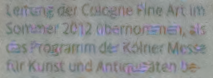}}
\end{minipage}
\hfill
\begin{minipage}[b]{\linewidth}
  \centering
  \centerline{\includegraphics[width=8.5cm]{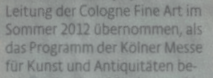}}
\end{minipage}
\caption{Newsprint (top), captured by a smartphone Quad Bayer CFA,
  converted to Bayer by the on-sensor remosaicker and then demosaicked
  by the state of the art algorithm of \cite{monno2017adaptive}. The
  same newsprint (bottom) captured by an equivalent sensor with the
  RGBW CFA and demosaicked with the algorithm of \cite{singh2011icip}.
  Note the poor legibility and severe artifacting in the Quad Bayer
  image.}
\label{fig:cameraCaptures}
\end{figure}

\begin{figure*}
\begin{tabular}{ccc}
\subfloat[Ground Truth]{\includegraphics[width = 5.6cm]{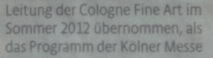}} &  
\subfloat[Hexadeca Bayer]{\includegraphics[width = 5.6cm]{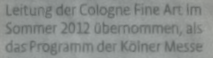}} &
\subfloat[RGBW-IA]{\includegraphics[width = 5.6cm]{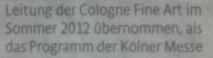}}\\
\subfloat[Quad-RGBW]{\includegraphics[width = 5.6cm]{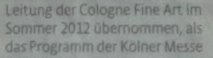}} &
\subfloat[LMS]{\includegraphics[width = 5.6cm]{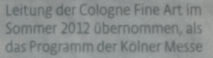}} &
\subfloat[Quad-LMS]{\includegraphics[width = 5.6cm]{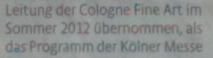}}\\
\end{tabular}
\caption{Simulated mosaicking of the Newsprint image with the Hexadeca
  Bayer and proposed CFAs followed by demosaicking with the algorithm
  of \cite{singh2011icip}.}
\label{fig:simulated}
\end{figure*}

\begin{table*}[b]
\begin{center}
  \begin{tabular}{| l | c | c | c | c | c |} 
  \hline
      {\bf Color Filter} & {\bf Bin} & {\bf Low Light } & {\bf Bright Light} &{\bf Power} & {\bf Frame} \\ 
  {\bf Array} & {\bf Mode} & {\bf SNR Advantage} & {\bf SNR Advantage} & {\bf Consumption} & {\bf Rate}\\ \Xhline{3\arrayrulewidth}
  {\bf Quad Bayer} & full resolution & 0 dB & 0 dB &{\bf 100\%} & {\bf 1x} \\ 
  & & (reference) & (reference) & (reference) &  (reference) \\ \cline{2-6}
  & binned once & 12 dB & 6 dB &{\bf 25}\% & {\bf 4x}\\ \Xhline{3\arrayrulewidth}
  {\bf RGBW-Kodak} & full resolution & {\bf 6} dB & {\bf 3} dB & {\bf 100\%} & {\bf 1x}\\ \cline{2-6}
  & binned once & {\bf 15} dB & {\bf 9} dB & 50\% & 2x \\ \Xhline{3\arrayrulewidth}
  {\bf RGBW-IA} & full resolution & {\bf 6} dB & {\bf 3} dB &{\bf 100\%} & {\bf 1x}\\ \cline{2-6}
  & binned once & {\bf 15} dB & {\bf 9} dB & 50\% & 2x\\ \Xhline{3\arrayrulewidth}
  {\bf LMS} & full resolution & 4.2 dB & 2.1 dB &{\bf 100\%} & {\bf 1x}\\ \cline{2-6}
  & binned once & 13.2 dB & 8.1 dB & 50\% & 4x\\ \Xhline{3\arrayrulewidth}
  \end{tabular}
\end{center}
\caption{Single Binnable Color Filter Array Summary}
\label{tab:singlesummary}
\end{table*}

\begin{table*}
\begin{center}
  \begin{tabular}{| l | c | c | c | c | c | c |} 
  \hline
      {\bf Color Filter} & {\bf Bin} & {\bf Low Light } & {\bf Bright Light} &{\bf Power} & {\bf Frame} \\ 
  {\bf Array} & {\bf Mode} & {\bf SNR Advantage} & {\bf SNR Advantage} & {\bf Consumption} & {\bf Rate}\\ \Xhline{3\arrayrulewidth}
  & full resolution & 0 dB & 0 dB & {\bf 100\%} & {\bf 1x} \\
  & & (reference) & (reference) & (reference) &  (reference) \\ \cline{2-6}    
  {\bf Hexadeca Bayer} & binned once & 12 dB & 6 dB & {\bf 25\%} & {\bf 4x} \\ \cline{2-6}
  & binned twice & 18 dB & 12 dB & {\bf 6.25\%} & {\bf 16x} \\ \Xhline{3\arrayrulewidth}
  & full resolution & {\bf 6} dB & {\bf 3} dB & {\bf 100\%} & {\bf 1x} \\ \cline{2-6}
  {\bf Quad-RGBW} & binned once & {\bf 18} dB & {\bf 9} dB & {\bf 25\%} & {\bf 4x} \\ \cline{2-6}
  & binned twice & {\bf 24} dB & {\bf 15 dB} & 12.5\% & 8x\\ \Xhline{3\arrayrulewidth}
  & full resolution & 4.2 dB & 2.1 dB & {\bf 100\%} & {\bf 1x} \\ \cline{2-6}
  {\bf Quad-LMS} & binned once & 16.2 dB & 8.1 dB & {\bf 25\%} & {\bf 4x} \\ \cline{2-6}
  & binned twice & 22.1 dB & 14.1 dB & 12.5\% & 8x\\ \Xhline{3\arrayrulewidth}  
  \end{tabular}
\end{center}
\caption{Double Binnable Color Filter Array Summary}
\label{tab:doublesummary}
\end{table*}

Figure \ref{fig:simulated} shows a crop of the aforementioned
newsprint image after mosaicking with both single and double binnable
CFAs followed by demosaicking with the method of \cite{singh2011icip}.
Figure \ref{fig:cameraCaptures} shows crops of the same newsprint
image captured by actual smartphone camera modules with Quad Bayer and
RGBW-Kodak CFA patterns. The Quad Bayer raw data was translated to
Bayer raw data by the on-sensor remosaicker followed by Bayer
demosaicking using the state-of-the art algorithm of
\cite{monno2017adaptive}. The RGBW-Kodak raw data was demosaicked by
the method of \cite{singh2011icip}.

Note the poor legibility of the Quad Bayer image in Figure
\ref{fig:cameraCaptures}, despite several generations of commercial
algorithm development. Its poor performance is also reflected in its
lower PSNR. Other binnable CFAs are free of such problems and yield
higher PSNRs.

\section{Conclusion}

We analyzed the Quad Bayer CFA and revealed a degeneracy in its
spectrum that limits its demosaicking quality. This issue was
illustrated by capturing a newsprint image with a commercial
smartphone sensor employing the Quad Bayer pattern. Furthermore, we
examined the upcoming Nona and Hexadeca Bayer CFAs and demonstrated
their freedom from the problem faced by the Quad Bayer.

Our analysis extended to the less popular RGBW-Kodak CFA, where we
identified a spectral issue leading to confusion between luminance and
chrominance signals on diagonal image features. Additionally, we noted
that its 2:1 floating diffusion binning reduced its binned-mode SNR
advantage over the 4:1 binned Bayer from 6 dB to 3 dB.

Finally, we introduced a pair of single and double binnable RGBW CFAs,
as well as a pair of single and double binnable LMS CFAs. We
demonstrated that their spectra are free from issues and
experimentally verified their good demosaicking performance. Both
double binnable CFAs are composed of 2x2 tiles of like-colored pixels,
enabling 4:1 floating diffusion binning, leading to high SNR.  The
single binnable CFAs are summarized in table \ref{tab:singlesummary}
and the double binnable CFAs are summarized in table
\ref{tab:doublesummary}.

\vfill\pagebreak

\bibliographystyle{IEEEbib}
\bibliography{imaging}

\end{document}